\documentclass[twoside]{LCWS11}
\usepackage[latin1]{inputenc}
\usepackage[dvips]{graphicx,epsfig,color}
\usepackage{wrapfig,rotating}
\usepackage{amssymb,amsmath,array}
\usepackage{cite}
\begin{document}
\title{
Muon Background Reduction in CLIC} 
\author{L.~C.~Deacon$^1$, G.~Blair$^2$ and H.~Burkhardt$^1$
\vspace{.3cm}\\
1- CERN\\
Meyrin, Geneva - Switzerland
\vspace{.1cm}\\
2- Royal Holloway, University of London \\
Egham Hill, Egham, Surrey - United Kingdom\\
}

\maketitle

\begin{abstract}
We report on a study concerning the reduction of muon backgrounds in CLIC using magnetised iron. 
\end{abstract}

\section{Introduction}
We previously reported on a study of muon backgrounds in CLIC \cite{Burkhardt:2010zzb}. Subsequently, the CLIC detector group reported that such muon fluxes would result in a muon background which would critically degrade the performance of the detector \cite{Linssen:2012hp}. Here, we report on a study of how introducing magnetised toroids in the beam delivery system would affect the flux of muons hitting the detector. As in \cite{Burkhardt:2010zzb}, the tracking was done in two stages: firstly, a halo distribution was generated using HTGEN-PLACET \cite{htgen}, \cite{Latina:2008zz}. This was then tracked from the first vertical betatron spoiler to the detector using BDSIM \cite{Agapov:2009zz}, including electromagnetic and muon production processes in the simulation.

\section{Simulation procedure}
\subsection{Halo generation and tracking}
The types of processes contributing to halo formation \cite{haloProcesses} include particle processes and optics related processes. The particle processes contributing to halo formation include beam-gas scattering (both elastic and inelastic) and thermal photon scattering. Estimates of the effects of these processes exist. Optics-related processes include mismatch, coupling, dispersion and non-linearities. These would require tracking with a realistic machine. Tracking has so far been done with a perfect machine only. Various other processes exist such as noise and vibrations, dark current and wakefields, whose effect on the amount of halo has not yet been calculated. By experience, the amount of actual halo is very hard to predict and may vary considerably in a given machine.

We assume that the beams are cleaned before entering the linac so we only need to consider the extra halo produced in the linac and BDS. The HTGEN code was used to generate halo by beam-gas scattering. (Mott scattering) and inelastic scattering (bremsstrahlung). HTGEN is interfaced with PLACET, which allows halo tracking alongside tracking the core of the beam. 

The 2007 estimate of the linac beam gas pressure was 10 nTorr. In light of the results presented in \cite{Burkhardt:2010zzb}, and because of requirements related to beam stability, this was updated to 1 nTorr. Together with a larger emittance, this reduced the total linac probability of beam gas scattering by a factor of 13, so that we would now expect $1.5 \times 10^{-5}$ of the electrons to be lost in the spoilers \footnote{A recent analysis seems to indicate that the fraction of electrons hitting the collimators is significantly lower than the estimates presented here. The muon flux to the detector decreases proportionally.}. Work is ongoing to update and improve the halo simulations.

\subsection{BDSIM}
BDSIM \cite{Agapov:2009zz} is a toolkit based on Geant4 \cite{Agostinelli:2002hh}, thus giving access to many electromagnetic and hadronic interaction models as well as a powerful geometry description framework. On top of this, fast particle tracking routines and some additional physics processes are introduced, and a high level geometry description language is added. An interface to PLACET has also been developed in order to combine PLACET's wake-field effects on the beam halo, with BDSIM's capability for secondary generation and tracking.  Details of the simulated geometry and physics processes are given in \cite{Burkhardt:2010zzb}

\subsection{Muon spoilers}\label{S:muonSpoil}

Muon spoilers were introduced in the simulation. The somewhat simplified geometry used in the simulation consists of iron cylinders with an outer radius of 55~cm, a 1~cm inner radius, with a 1.5 Tesla solenoid field within the cylinder. The muon spoilers are placed ~100m downstream of each of the four sets of betatron collimators, to allow the muon distribution to expand before hitting the spoilers, and have a total length of ~80m. The details of the lengths and locations are given in table \ref{tab:muSpoil}.  

\begin{wraptable}{l}{0.5\columnwidth}
\centerline{\begin{tabular}{|l|r|r|}
\hline
Muon spoiler  & $Z$~[m]   &   Length~[m] \\\hline
    MS1a        & 99      &  8   \\
    MS1b        & 117     &  8      \\
    MS2a        & 211     &  9      \\
    MS2b        & 229     &  9      \\
    MS3a        & 323     &  11     \\
    MS3b        & 341     &  11     \\
    MS4a        & 406     &  14     \\
    MS4b        & 435     &  14     \\
\hline
\end{tabular}}
\caption{Muon spoiler lengths and locations. Here $Z$ is the distance along the reference trajectory from the start of the first vertical spoiler YSP1 to the start of the muon spoiler.}
\label{tab:muSpoil}
\end{wraptable}

\section{Muon estimates}
In \cite{Burkhardt:2010zzb} we reported an estimated muon flux of $2\times 10^{6}$ muons per train, many of which would be seen as background in the detector, and that reducing the muon flux would require magnetised shielding. In the full simulation, the number of muons per bunch hitting the detector was reported as $207 \pm 2$.  

The CLIC detector CDR states that at the level of 1 muon per bunch crossing does not constitute an important problem for the detector \cite{ Linssen:2012hp}. A factor 13 improvement is given by reducing the linac beam gas pressure from 10 nTorr to 1 nTorr. Further improvement would be needed to reduce the muon flux to an acceptable level. 

With the muon spoilers added, as described in \ref{S:muonSpoil}, the flux is reduced by a further factor of 10. 

\section{Optimisation}
\subsection{First Iteration}
\begin{figure}
\centerline{\includegraphics[width=0.45\columnwidth]{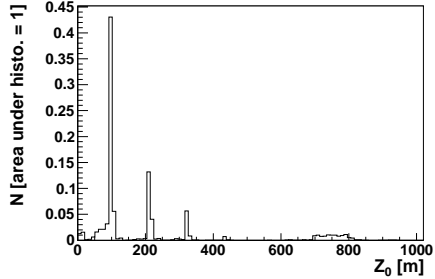}}
\caption{Origin or location of last scatter of muons hitting the detector (within 6~m radius at the final focus quadrupole) with no muon spoilers. The area under the histogram is normalised to 1.}\label{F:z0_0}
\end{figure}
To further reduce the muon flux to the detector, the origins of the remaining muons were analysed. A histogram of $Z_{0}$ was plotted, Figure \ref{F:z0_0}, where $Z_{0}$ is the location at which a muon was created, for all muons reaching the detector, in terms of distance from the first muon spoiler YSP1 along the reference particle trajectory. Four peaks can be seen, which correspond to the locations of the betatron collimators. There is an additional, wider peak corresponding to a long dipole, from $\sim 600 - \sim 800$~m, which is attributed to muons hitting this dipole and producing further particles. These results suggested that the addition of a magnetised tunnel filler at $\sim 625$~m  downstream of the first vertical betatron spoiler could help reduce the muon background further.

\begin{figure}
\centerline{\includegraphics[width=0.45\columnwidth]{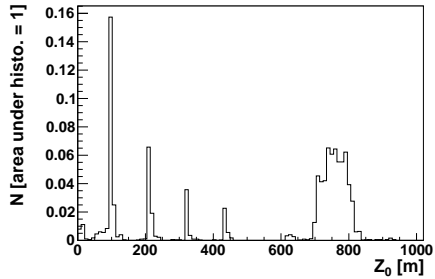}}
\caption{As Figure \ref{F:z0_0}, but with muon spoilers. The area under the histogram is normalised to 1.}\label{Fig:z0_2}
\end{figure}

\subsection{Second Iteration}
\begin{figure}
\centerline{\includegraphics[width=0.45\columnwidth]{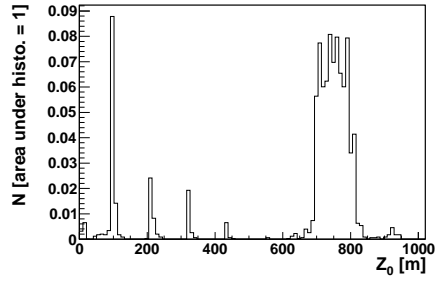}}
\caption{As Figure \ref{F:z0_0}, but with muon spoilers and tunnel filler. The area under the histogram is normalised to 1.}\label{Fig:z0_5}
\end{figure}

An outer radius 3.5~m, inner radius 1~cm, 17~m long tunnel filler was added to the simulation starting at $s=644$~m. This simulation showed that the addition of the tunnel filler would improve the muon flux by a further factor of 1.5, and its $Z_{0}$ histogram, figure \ref{Fig:z0_5}, suggested that the muons seemed to be generated in the long dipole section.
 
\subsection{Third iteration}

\begin{figure}
\centerline{\includegraphics[width=0.45\columnwidth]{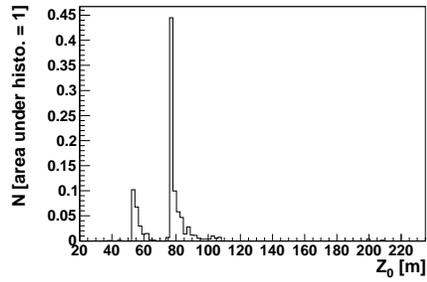}}
\caption{Origins of muons created in the final focus dipole by secondary electrons, positrons and photons from the beam delivery system collimation system. $Z_{0}=0$ corresponds to the start of final focus dipole magnet BFF3. The area under the histogram is normalised to 1.}\label{F:originsbff3it3}
\end{figure}

All available drift spaces were filled with magnetised tunnel fillers - 3.5~m outer radius - and the flux was found to be similar to the second iteration. We conclude that adding more magnetised iron beyond that in the second iteration will probably not reduce the muon flux any further. 90\% of the remaining muons are created in the long dipole section. Particles inside the beam pipe were tracked from the start of final focus dipole BFF3. The origin of muons hitting the detector in distance from the start of BFF3 (not normalised) is shown in Fig. \ref{F:originsbff3it3}. From left to right, the three peaks correspond to the impacts of positrons, photons and electrons, respectively. 
\nolinebreak

The positrons contribute $\sim 20\%$ to the muon flux, the photons $\sim 75\%$ and the electrons $\sim 5\%$. These secondary particles were originally created upstream in the apertures of the beam delivery system. There is not enough space in which to deflect the muons created by these secondaries in the final dipole.

\section{Conclusion}
Work is ongoing to provide a new set of HTGEN-PLACET simulations of the linac-BDS system. These will provide better estimates of the halo to input into the BDSIM simulation. There is a predicted factor 13 reduction in the halo after reducing the beam gas pressure in the linac from 10 nTorr to 1 nTorr and a larger emittance. With the addition of the muon spoilers outlined above, plus the tunnel filler, the muon flux is reduced by a factor of 15. With the muon spoilers alone, this is a factor of 10. These reductions combine to give a predicted 1.5 muons per beam per bunch crossing. Most of the remaining muons are created in the final focus dipoles from secondary positrons and photons from further upstream - these could possibly be reduced by improving the collimation system. With further improvements, a muon flux of 1 per bunch crossing, requested by the detector, seems feasible. Further optimisation of the muon spoiler system may be needed.


\begin{footnotesize}
\bibliographystyle{unsrt} 
\bibliography{lcws11_ld}
\end{footnotesize}


\end{document}